\documentclass{emulateapj}

\usepackage[]{graphicx}

\shorttitle{RELATIONSHIP BETWEEN STAR FORMATION RATE AND BLACK HOLE ACCRETION}
\shortauthors{Rodighiero, Brusa, Daddi et al.}

\begin{document}

\title{RELATIONSHIP BETWEEN STAR FORMATION RATE AND BLACK HOLE ACCRETION AT \lowercase{z}=2:\\
 THE DIFFERENT  CONTRIBUTIONS IN  QUIESCENT, NORMAL AND STARBURST  GALAXIES }

\author{
G. Rodighiero\altaffilmark{1,}\altaffilmark{2},
M. Brusa\altaffilmark{3,5,7},
E. Daddi\altaffilmark{4},
M. Negrello\altaffilmark{6},
J. R. Mullaney\altaffilmark{8},
I. Delvecchio\altaffilmark{3},
D. Lutz\altaffilmark{7},
A. Renzini\altaffilmark{6},
A. Franceschini\altaffilmark{1},
I. Baronchelli\altaffilmark{1},
F. Pozzi\altaffilmark{3},
C. Gruppioni\altaffilmark{5},
V. Strazzullo\altaffilmark{4,9},
A. Cimatti\altaffilmark{3},
J. Silverman\altaffilmark{10}
}

\altaffiltext{1}{Dipartimento di Fisica e Astronomia "G. Galilei", Universita' di Padova, Vicolo dell'Osservatorio 3, I-35122,  Italy}
\altaffiltext{2}{email: giulia.rodighiero@unipd.it}
\altaffiltext{3}{ Dipartimento di Fisica e Astronomia, Universit\`a di Bologna,  viale Berti Pichat 6/2, 40127 Bologna, Italy}
\altaffiltext{4}{Laboratoire AIM, CEA/DSM-CNRS-Universit{\'e} Paris Diderot, IRFU/Service d'Astrophysique, B\^at.709, CEA-Saclay, 91191 Gif-sur-Yvette Cedex, France.}
\altaffiltext{5}{INAF-Osservatorio Astronomico di Bologna, Via Ranzani 1, I-40127, Bologna, Italy}
\altaffiltext{6}{INAF-Osservatorio Astronomico di Padova, Vicolo dell'Osservatorio 2, I-35122 Padova, Italy}
\altaffiltext{7}{Max Planck Institut f\"ur Extraterrestrische Physik, Giessenbachstrasse 1, 85748 Garching bei M\"unchen, Germany}
\altaffiltext{8}{Department of Physics and Astronomy, University of Sheffield, S3 7RH, U.K.}
\altaffiltext{9}{Department of Physics, Ludwig-Maximilians-Universit{\"a}t, Scheinerstr. 1, 81679 M{\"u}nchen, Germany}
\altaffiltext{10}{Kavli Institute for the Physics and Mathematics of the Universe (WPI), Todai Institutes for Advanced Study, The University of Tokyo, Kashiwanoha, Kashiwa 277-8583, Japan. }

\begin{abstract}
We investigate the co-evolution of black-hole-accretion-rate (BHAR) and star-formation-rate (SFR) in $1.5<z<2.5$ galaxies displaying a greater diversity of star-forming properties compared to previous studies.  We combine X-ray stacking and far-IR photometry of stellar mass-limited samples of normal star-forming, starburst and quiescent/quenched galaxies in the COSMOS field.  We corroborate the existence of a strong correlation between BHAR (i.e. the X-ray luminosity, $L_X$),
and stellar mass ($M_*$) for normal star-forming galaxies, although find a steeper relation than previously reported.  We find that starbursts show a factor of 3 enhancement in BHAR compared to normal SF galaxies (against a factor of 6 excess in SFR), while quiescents show a deficit of a factor $\times5.5$ at a given mass.  One possible interpretation of this is that the starburst phase does not coincide with cosmologically relevant BH growth, or that starburst-inducing mergers are more efficient at boosting SFR than BHAR.  Contrary to studies based on smaller samples, we find the BHAR/SFR ratio of main sequence (MS) galaxies is not mass invariant, but scales weakly as $M_*^{0.43\pm0.09}$, implying faster BH growth in more massive galaxies at $z\sim2$.  Furthermore, BHAR/SFR during the starburst is a factor of 2 lower than in MS galaxies, at odds with the predictions of hydrodynamical simulations of merger galaxies that foresee a sudden enhancement of $L_X/$SFR during the merger.  Finally, we estimate that the bulk of the accretion density of the Universe at $z\sim2$ is associated with normal star-forming systems, with only $\sim6(\pm1)\%$ and $\sim11(\pm1)\%$ associated with starburst and quiescent galaxies, respectively.

\end{abstract}


\section{Introduction}
The self-regulation of host galaxies and their central supermassive black holes (BH) is often invoked to explain the local scaling relations between BH mass and galaxy properties, 
(such as the stellar mass or the velocity dispersion of the bulge; Magorrian et al. 1998, Ferrarese \& Merritt 2000; Aller \& Richstone 2007).  As a consequence, determining the relative roles of the various processes that drive the co-evolution of BHs and galaxies has emerged as a key goal of current astrophysics research.

While the role of violent mergers is still considered as an important triggering mechanism for BH growth (observed as active galactic nuclei; AGN), in particular in very luminous quasars (Hopkins et al. 2008, Treister et al. 2012), the emerging observational framework supports the idea that secular processes are responsible for both star formation and SMBH growth in a large majority of galaxies displaying moderate nuclear activity (Cisternas et al. 2011, Kocevski et al. 2012).  As such, it is now clear that processes other than galaxy mergers, such as gas turbulence or disk instabilities, play an important role in triggering BH accretion (Bournaud et al. 2012).  However, with many studies finding no clear correlation between instantaneous BH growth and star formation (SF; Silverman et al. 2009, Shao et al. 2010, Rosario et al. 2012, Mullaney et al. 2012a, Azadi et al. 2014), actually measuring the contribution of these different triggering processes to the BH growth budget has proven a significant challenge.

A key step forward in this area has been achieved by studies that have adopted X-ray stacking analyses to derive the {\it average} BH growth rate of large samples of galaxies (i.e., including undetected, low accretion rate BHs). In doing so, these studies have demonstrated that the average BHAR is tightly correlated with both $M_*$ and SFR (Mullaney et al. 2012b, hereafter M12; Chen et al. 2013) and mimics the SFR-$M_*$ relation of normal star-forming galaxies (Elbaz et al. 2007, Daddi et al. 2007, Rodighiero et al. 2014).  However, these previous studies did not discriminate between different types of galaxies, whereas it is known that different levels of SFR are triggered by different events 
which may also trigger different levels of BH growth.  Moreover, these studies did not consider BHAR in quiescent galaxies, leaving their contribution unexplored.

To constrain the relative BHAR in galaxies displaying different levels of star-forming properties (and thus, in different evolutionary stages), we employ X-ray stacking to investigate how the average X-ray luminosity (a proxy for BH growth rate) varies across bins of stellar mass and specific SFR (sSFR)
using large statistical stellar mass-selected samples.  Exploiting the unique depth/area combination of the COSMOS field (Scoville et al. 2007) means we avoid the small number statistics that prevented previous X-ray stacking studies to seperate their samples in terms of sSFR.  Here, we focus on the $1.5<z<2.5$ epoch when most of today's stars were formed; 
see also Delvecchio et al. (2015, subm.).

Throughout the paper we use a Chabrier (2003) initial mass function and $H_0=71$ km s$^{-1}$, $\Omega_{\Lambda}=0.73$, $\Omega_{M}=0.27$.

\section{Splitting the SFR-M$_*$ plane at  $1.5<\lowercase{z}<2.5$}
\label{selection}

\subsection{Normal star-forming galaxies}
\label{MS}
We use the K-band selected sample of $1.5< z<2.5$ star-forming galaxies down to K$_{s,AB}< 23$ in the COSMOS field (McCracken et al. 2010) selected according to the BzK criterion designed to select star-forming galaxies at these redshifts (the star-forming BzK, hereafter sBzK; Daddi et al. 2004).  Stellar masses have been computed following the same procedure as in Daddi et al. (2007).  We have considered the potential issue of AGN contamination in the computation of stellar masses (in particular in the highest mass bin): we verified that the AGN stellar masses computed with a spectral decomposition that accounts for a dusty torus emission, provides on average consistent results with $M_*$ classically derived by a pure stellar component (no offset and a dispersion of $\sim$0.25 dex for both type 1 and type 2 AGN).
SFRs of these sBzK galaxies are estimated from the UV rest-frame luminosity corrected for dust extinction, with reddening being inferred from the slope of the UV continuum as in Daddi et al. (2007).  However, to factor-in the potential contamination from the type 1 AGN to the UV emission, we will use the average SFR(IR+UV) derived from stacking the sBzK sources on the PACS/{\it Herschel} maps in different mass bins (as in Rodighiero et al. 2014).

Limiting our analysis to M$_*>$10$^{10}$M$_\odot$, where the sample is complete (see Ilbert et al. 2010), results in a final sBzK sample of 16,435 galaxies. See black points in Figure 1, top panel.

\begin{figure}
\centering
\includegraphics[width=8.5cm]{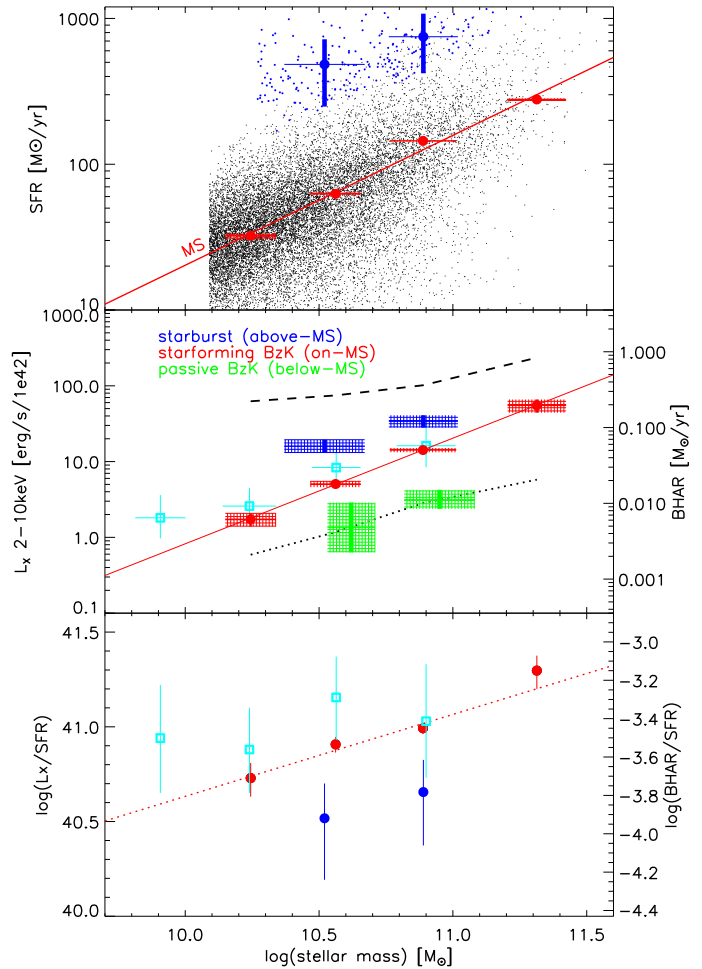} 
\caption{ {\it Top panel}: The mass-SFR relation.  We report the sBzK sample with their SFR(UV) (black points), while the corresponding average SFR(IR+UV) derived from the PACS stacking analysis are reported as red filled circles (Rodighiero et al. 2014).  
The linear fit provides an MS relation as log[$SFR$]=$0.89(\pm0.09)\times$log[M$_*$]+14.02.
We also show the starburst sample (blue points) with the corresponding average SFR(IR+UV) in the two mass bins (errors on average SFR are the standard deviation of the SFR distribution within each mass bin).  {\it Middle panel}: stellar mass--average $L_X$ relation at $1.5<z<2.5$ from our X-ray stacking analysis.  Shaded colour regions are the results for the three samples: starburst (blue), main sequence (red) and quiescent sources (green).  We provide a fit for the MS sample (solid red line, log[$L_X$]=$1.40(\pm0.09)\times$log[M$_*$]-14.11).  
The open cyan squares are the results of M12 in GOODS-S, scaled to the same K-correction adopted in this work.  The dashed and dotted black lines represent the average $L_X$ of detected and stacked X-ray sources in the MS, respectively.  {\it Bottom panel}: $L_X$/SFR relation for MS (red) and starburst (blue). The dotted line is a linear fit to the relation for the MS sample (log[$L_X$/SFR]=$0.43(\pm0.09)\times$log[M$_*$]+35.6) and the open cyan squares are again the results of M12.
In each panel the width of the stacked data along the x--axis represents the RMS of the mass distribution in each bin.}
\label{fig1}
\end{figure}

\subsection{Starburst sources} 
\label{offMS}
For our starburst sample, we start from the Gruppioni et al. (2013) sample selected from PACS/{\it Herschel} PEP observations in the COSMOS field (Lutz et al. 2011) at 100$\mu$m and 160$\mu$m.
We refer to Berta et al. (2010) and Gruppioni et al. (2013) for a description of the data catalogues.  Stellar masses have been derived by fitting the broad-band SEDs of our sources using a modified version of MAGPHYS (Da Cunha et al.  2008, Berta et al. 2013).  
Such $M_*$ are consistent with those derived with the formalism of Daddi et al. (2007, zero offset and a dispersion of $\sim$0.3dex).
The SFRs of these galaxies are derived from the total IR luminosities. 

To classify a sample of starburst galaxies lying above the MS, we have applied the same criterion as Rodighiero et al. (2011, i.e., sSFR 0.6 dex above the MS) over the whole $1.5<z<2.5$  range.  This selection provides a total number of 231 candidate starbursts with log(M$_*$)$>10.27$ M$_\odot$.  See blue points in Figure 1, top panel.

\subsection{Quiescent galaxies} 
\label{pBzK}
To complete the study of galaxies populating the SFR-M$_{*}$ plane, 
we include the complementary sample of passive BzK (pBzK). 
A description of the pBzK sample (including photo-$z$) in the COSMOS field is presented in Strazzullo et al. (A\&A submitted).   
$M_*$ have been derived as for sBzK parent sample and SFR derived via SED fitting are on average consistent with zero. However, as well as purely quenched, red and dead galaxies, the pBzK sample also consists of galaxies with low SFR.  
The adopted selection provides a total of 2559 quiescent sources with log(M$_*$)$>10.4$ M$_\odot$.

\section{X-ray data}
\label{X-ray}
The central 0.9 deg$^{2}$ of the COSMOS field has been imaged by {\it Chandra} for a total exposure time of 1.8 Ms, down to a depth of $\sim2\times10^{-16}$ erg cm$^{-2}$ s$^{-1}$ in the soft band (Puccetti et al. 2009).  A main catalog of 1761 X-ray point sources detected in at least one band (0.5-2, 2-8, or 0.5-8 keV) has been published by Elvis et al. (2009), together with optical identifications, optical magnitudes and spectroscopic and photometric redshifts from different spectroscopic campaigns and/or from SED fitting estimates (Civano et al. 2012).

We cross-correlated the different samples described above with the Civano et al. (2012) counterparts catalog (using a matching radius of 1"). We considered only objects falling in the Chandra area ($\sim50\%$ of the total samples), recovering a total of 317, 19 and 47 matches for the sBzK, PEP and pBzK samples, respectively.  The C-COSMOS exposure can be considered complete down to L(0.5-8 keV)=3$\times10^{43}$ erg s$^{-1}$ in the entire redshift range explored in this work.  As such, these X-ray detections probe the moderate to high luminosity tail of the AGN population ($L_X$$>10^{43.5-44}$ erg s${^{-1}}$), complementing the deeper exposures in the GOODS fields (see M12).

\begin{figure}
\centering
\includegraphics[width=9cm]{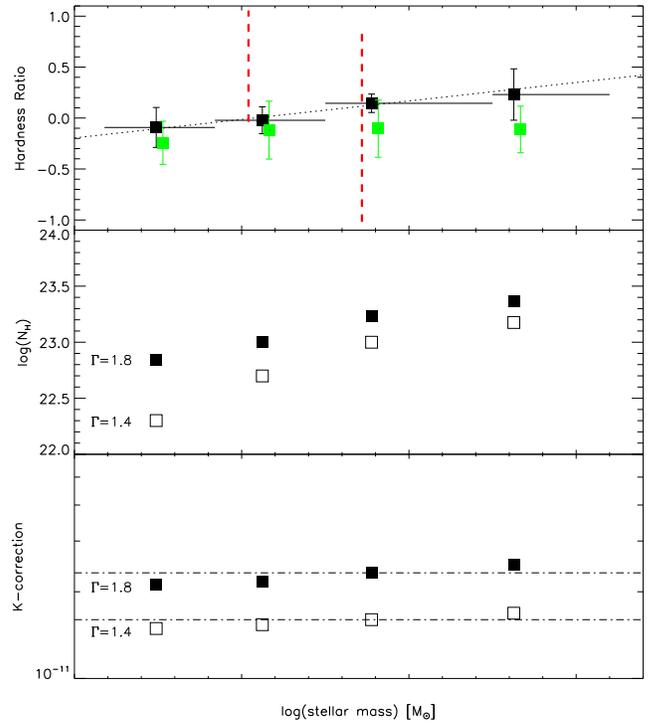}
\caption{{\it Top panel}: Stellar mass -- HR relation at $1.5<z<2.5$ for the MS-sBzK  stacked sample (filled black squares). The dotted line is a linear fit to the distribution (HR=0.31$\times$log[M$_*$]-3.3).
The vertical red dashed lines show the marginal constraints that we can derive for the starbursts, basically consistent with the MS.
The values for the detected sources are also reported (green squares).
{\it Middle panel}: Derived N$_H$ values for two different assumptions of the X-ray spectral slope $\Gamma$.
{\it Bottom panel}:  Final K-corrections for the two different values of $\Gamma$ turn to be almost constant with mass.
We adopt a constant value of $\Gamma$=1.4 and N$_H$=22. 
}
\label{fig2}
\end{figure}

\subsection{Stacked and average fluxes}
\label{stack_X}
In order to study the average X-ray properties of sources not individually detected at the limit of the C-COSMOS image, we used the CSTACK tool\footnote{http://cstack.ucsd.edu/cstack/} (v3.1, T. Miyaji).  To reduce to a minimum the contamination of detected sources in the stacked signal, we considered those galaxies without a known X-ray source within 10" from the centroid. This led to the exclusion in the analysis of about 5-10\% of sources in the various mass bins.  The total sample of stacked objects overlapping the {\it Chandra} area consists of 6393, 73 and 1008 objects in the sBzK, PEP and pBzK samples, respectively, at $M_*>10^{10} M_\odot$. The numbers of galaxies in each mass bin are reported in Table \ref{table}.  The count rates in the 0.5-2 (soft, S) and 2-7 keV (hard, H) bands are computed by averaging the background subtracted count-rates measured at each input position in the stacked sample (see Cimatti et al. 2013).  The CSTACK tool provides the results of a bootstrap re-sampling analysis, which uses a large number (500) of re-sampled stacked count rate. We take the median of the distribution as the best estimate of the stacked count rate, and assume the 5th to 95th percentile range as the associated confidence interval.

From the ratio of the soft and hard bands we derived the Hardness Ratio, HR=(H-S)/(H+S). The values of HR are the average of the distribution of the values derived by resampling (thousands of times) the distribution of the corresponding stacked count rate.
The HR for the stacked sBzK sample as a function of mass are shown in the upper panel of Figure 2. Although a trend of increasing HR with mass can be seen, the relation is also consistent with being flat. We therefore applied the same conversion factors from a single power-law in all mass bins to derive fluxes and luminosities.
An average value of HR$\sim$0 (hence N$_H\sim23$)  is similar to results of stacked studies of much smaller samples of undetected galaxies in the M*-SFR plane  (e.g. Olsen et al. 2013 reported an HR$\sim$0.09 for their SF sample).

When combining the detected and the stacked X-ray signals to calculate the total average X-ray flux (see middle panel of Fig.1) we first resample the corresponding distributions of flux values (assuming a Gaussian error distribution for the detected fluxes) and then derive the distribution of averaged X-ray fluxes using:
\begin{equation}
F_{\rm ave} = \frac{\sum_{i=1}^{n_{\rm detected}}F_{i, \rm detected} +  n_{\rm stacked}*F_{\rm stacked}}{n_{\rm detected} + n_{\rm stacked}},
\end{equation}
where $n_{\rm detected}$ and $n_{\rm stacked}$ are the numbers of X-ray detected sources and of stacked sources, respectively, while $F_{i, \rm detected}$ (with $i=1,n_{\rm detected}$) and $F_{\rm stacked}$ are the corresponding X-ray fluxes.
As above, we take as our best estimate of the combined stacked+detected X-ray flux the median of the distribution of the $F_{\rm ave}$ values with a 95\% confidence interval derived from the same distribution. When the stacked signal alone is not significant (i.e. consistent with 0 within 2$\sigma$, as for the starburst and quiescent samples) we assume as a measure of the lower and upper limits of the stacked+detected X-ray fluxes (or luminosities) the corresponding 5th and 95th percentiles.
The main reason for resorting to stacking in our work is to estimate integrated SFR and $L_X$ outputs over well defined galaxy population, rather than typical $average$ quantities for each object in the sample.

 Absorption-corrected rest-frame 2-10 keV fluxes (and corresponding luminosities) are derived from the count rates 
  in the observed hard 2-7 keV band
 by assuming an absorbed power-law with $\Gamma$=1.4 with the relevant conversion factors (as derived by WEBPIMMS) and K-corrections. A different choice of $\Gamma$ (e.g. $\Gamma=1.8$) would provides a higher normalisation for $L_X$, but which is still consistent with the results obtained for $\Gamma=1.4$ and therefore would not change our main results.  Similarly, applying a mass-dependent K-correction derived from the HR with the same power-law and varying column density would only change $L_X$ by a maximum of 15\% (lower panels of Figure 2).  Table \ref{table} lists the 2-10 keV X-ray luminosities for all the stacked samples.

We have verified that the average contribution to the final (stacked+detected) $L_X$ from star-formation is always lower than 10\%, by using the SFR-$L_X$ relation of Vattakunnel et al. (2012), i.e. within the uncertainties of our data, and thus it does not affect our results.
We note that the fractional contribution of SFR to the stacked signal is much larger ($\sim$50\%), but it does not affect significantly $F_{\rm ave}$ in eq. 1.

\begin{figure} \includegraphics[width=9cm,height=12cm]{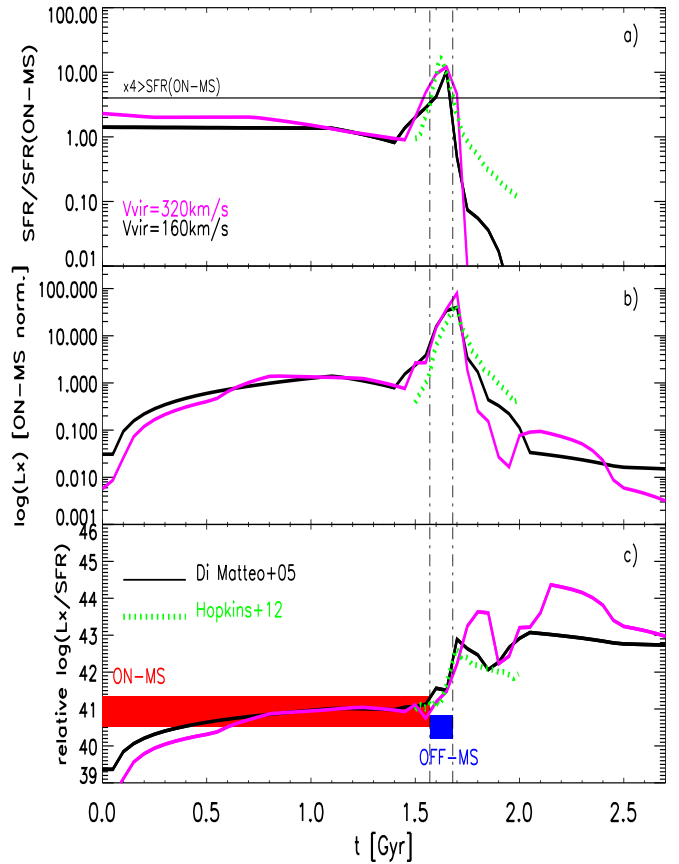} 
\caption{Comparison of our results with predictions of hydrodynamical models that simultaneously follow star formation and the growth of black holes during a galaxy--galaxy merger (Di Matteo et al. 2005, Hopkins et al. 2012).   {\bf a)} To define the simulated merger/starburst phase 
    we report the SFR history of the models normalized to the pre-merger phase. 
    The temporal window where (SFR)/(SFR pre-merger) is larger than a factor 4 for all models is marked by the two vertical dot-dashed lines.  {\bf b)} Evolution of the X-ray luminosity with time.  {\bf c)} Evolution of the X-ray luminosity to SFR ratio, as a function of time.  Two different values of the virial velocity of the collisional galaxies are considered in the Di Matteo et al. (2005) model ($v$=160km/s black lines, $v$=320km/s magenta lines).  Our results are presented as a red rectangle for the pre-merger phase (corresponding to our MS normal star-forming sources) and as a blue region for the merger phase (i.e. our starburst sample). The width of $L_X$/SFR from our data spans the full range covered in all mass bins (see Fig. \ref{fig1}, bottom panel).  } \label{fig3} \end{figure}

\section{Results and discussion}
The average $L_{X}$ (stack + detections) vs. $M_*$ for the MS, starburst and quiescent samples is shown in the middle panel of Figure 1 (with the relation derived separately for X-ray detected and stacked sources).  Our analysis shows a robust $L_{X}-M_*$ correlation (the hidden AGN MS of M12) for normal star-forming galaxies, confirming its existence also at higher $M_*$  This is a superlinear relation with a slope of 1.40($\pm0.09$) in logarithmic units.  The presence of this sequence, together with the SFR-$M_*$ MS, has been interpreted as a footprint of the same secular processes that feed and sustain both the SFR and the BHAR from the galaxies gas reservoirs.  We note, however, that the relation observed in this work is steeper than that reported by M12 (who found a slope of 1.05$\pm0.36$, consistent with their larger errors) at the same redshift.  In particular, our increased statistics at higher masses provides robust constraints on this slope. 
Interestingly, starburst galaxies show higher levels of BHAR, with respect to MS galaxies at fixed $M_*$.  By contrast, quiescent galaxies have a deficit of an average factor 5.5 in terms of $L_X$ with respect to the MS, suggesting that they belong to an evolutionary phase where the gas fuelling the central engine (and the SFR) is almost exhausted (or the accretion is simply temporarily suppressed).  Given the number densities and average $L_X$ of starburst, normal star-forming and quiescent galaxies in the mass bin $10.75<log(M_*)<11.25$ (where all samples are complete, spanning the X-ray luminosity range $41.8<log(L_X)<43.6$ ), we argue that the accretion activity during the starburst phase is just $\sim6(\pm1)\%$ of the cosmic integral BHAR at this redshift, while that of quiescent sources is $\sim11(\pm1)\%$.  The bulk of the accretion density ($\sim83(\pm1)\%$) of the Universe is then associated with normal star-forming systems (similarly to what is found at $z<1.2$, Georgakakis et al. 2014).


\begin{table*}
\begin{scriptsize}
\centering
Main statistical properties of the sample\\
\begin{tabular}{cccccc}
\hline
{\bf {Main Sequence}}& & & & &\\
\hline
(1) & (2) & (3) & (4) & (5) & (6) \\
{\it log(Mass range)} & $N_{stacked}$ & $N_{detections}$ & $<M_*>$ & $<SFR>$ & $<L_{2-10keV}>$\\
\hline
10.09--10.42 & 3215 &   56 & 10.24$\pm$0.09 &    37.1$\pm$7.5   &$1.73^{+0.35}_{-0.34}$ \\
10.42--10.75 & 2158 & 119 & 10.56 $\pm$0.09 &   65.3$\pm$13.1 &$5.06^{+0.48}_{-0.46}$ \\
10.75--11.25 &   984  & 124 & 10.88$\pm$0.12 & 149.3$\pm$29.9 &$14.20^{+0.94}_{-0.93}$\\
11.25--12.00 &    36   &   10 & 11.31$\pm$0.11 &  310.8$\pm$62.2  &$55.12^{+10.96}_{-10.96}$ \\
\hline
{\bf {Starbursts above MS}}& & & & &\\
\hline
{\it log(Mass range)} & $N_{stacked}$ & $N_{detections}$ & $<M_*>$ & $<SFR>$ & $L_{2-10keV}$\\
10.25--10.75 & 48 & 12 & $10.52\pm0.15$ & $484.0\pm235.4$ &13.0--19.4\\
10.75--11.25 & 25 &  6 & $10.89\pm0.13$ & $749.6\pm328.8$ &28.1--40.9\\
\hline
{\bf {Quiescent below MS}}& & & & &\\
\hline
{\it log(Mass range)} & $N_{stacked}$ & $N_{detections}$ & $<M_*>$ & $<SFR>$ & $L_{2-10keV}$\\
10.42--10.75 & 381 & 4   & $10.62\pm0.09$ & -- & 0.63--2.88 \\
10.75--11.25 & 627 & 20 & $10.95\pm0.13$ & -- &  2.37-4.23 \\
\hline
\end{tabular}
\caption {Notes: (1) $M_*$ range in solar units.
(2) Number of X-ray stacked sources.
(3) Number of sources detected in the Chandra X-ray catalogues.
(4) Average stellar mass, errors are the 1-$\sigma$ of the mass distribution in each bin ($log(M_\odot)$).
(5) Average SFR ($M_\odot/yr$).
(6)  Average intrinsic AGN rest-frame 2-10 keV X-ray luminosity ($10^{42} erg s^{-1} cm^{-2}$) obtained by combining stacked and detected X-ray sources. Where the single stacked X-ray flux is not detected, we report 2$\sigma$ upper and lower limits.
}
\label{table}
\end{scriptsize}
\end{table*}

During the starburst phase, where SFR is enhanced by a factor of $\sim$6 on average (Fig. 1, top panel), the average $L_X$  is also enhanced, but by a smaller amount (a factor of $\sim$3, Fig. 1 middle panel). 
The average $L_X$/SFR ratio during the starburst phase is then a factor of $\sim2$ lower than that of the normal SF population (in the common mass range, Fig.1 lower panel).  
Keeping in mind that $L_X$ and SFR are dominated by different populations,
one possible interpretation of this result could be a delay in the BHAR enhancement relative to that of the SFR, as suggested by Hopkins et al. (2012, see also Lapi et al. 2014). However, as discussed below, our observations are inconsistent with the predictions of this model, and, thus, a simpler argument is that mergers are more efficient at boosting SFR than BHAR.  On the other hand, Alexander et al. (2008) found a similar lagging between the SFR and the BHAR in a sample of X-ray-obscured Submillimeter Galaxies at  $z\sim2$.  Moreover, state of the art semi-analytic models, based on a merger-driven scheme for starbursts and AGN (Lamastra et al. 2013), correctly foresee the observed increase of $L_X$ at increasing distance from the MS (i.e. the starburstiness) but they predict the same enhancement of both $L_X$ and SFR.

Thanks to the improved statistics of our sample at higher masses, we found that the $L_X$/SFR ratio is not invariant with the stellar mass, but scales as log($L_X/SFR)\propto M_*^{0.43\pm0.09}$ (Fig. 1, lower panel).  By contrast, M12 reported a constant value for this ratio as a function of mass.  Since $L_X$ traces the BHAR ($\dot M_{BH}$), a constant $L_X$/SFR ratio translates into a $M_{BH}/M_*$ constant ratio ($M_{BH}\propto M_*$, see formalism in M12).  By integrating the superlinear relation reported here, we instead find that $M_{BH}\propto M_*^{1.5}$, implying a steepening of the $M_{BH}/M_*$ distribution at $z\sim2$ 
(in line with Kormendy \& Ho 2013, who find a mass dependence of $M_{BH}/M_{bulge}$ in the local Universe).
However, we note the $L_X$/SFR trend presented here is consistent to within the errors bars shown in M12 within the common mass range (see cyan points in the lower panel of Fig. 1), and it is also consistent within the observed scatter of the local $M_{BH}/M_*$ relation (see Lasker et al. 2014).  Our results also suggest that BHs accrete more mass with respect to their host galaxies at the high mass end, as also seen in a small sample of luminous, obscured QSOs at $z\sim2$ (Bongiorno et al. 2014).

The definition of starbursts in our selection is consistent with what is envisaged in hydrodynamical simulations of merger galaxies that include feedback from a  BH.  This is shown in the panel a) of Figure 3, where we report the SF history of the Di Matteo et al. (2005) and Hopkins et al. (2012) models, normalized to the pre-merger phase. 
These models predict a sudden enhancement of $L_{\rm X}$ (panel b) and $L_{\rm X}$/SFR (panel c) during the merger phase, up to a factor of 100.
Within our sample of starburst-, sSFR-selected galaxies we do not find an enhanced $L_{\rm X}$/SFR ratio, which is actually slightly sub-MS. Clearly, the mere fact that these galaxies are starbursting indicates that the current AGN activity is not able to suppress SF in these (merging) systems. On the other hand, the peak of BHAR may be delayed with respect to the peak  SFR (e.g., Chen et al. 2013, Hickox et al. 2014), taking place when the starburst has already subsided, which is however contrary to the above simulations, where starburst and rapid BH growth are simultaneous events.

We caution that among SF galaxies (for MS and in particular for the starbursts, Lanzuisi et al. 2014) we might have missed a fraction of Compton Thick AGN for which the adopted conversion factors to derive the intrinsic $L_X$ might be underestimated.  Surveys sensitive to more penetrating hard X-rays (e.g. from {\it NuSTAR}) are required to unravel this hidden extinction (Mullaney et al. in prep.).

\begin{acknowledgements}
GR acknowledges support from ASI (Herschel Science Contract  2011 - I/005/011/0).         
MB acknowledges support from the FP7 Career Integration Grant ``eEASy'' (CIG 321913). 
ED acknowledges funding support from ERC-StG grant UPGAL 240039 and ANR-08-JCJC-0008.
We thank the anonymous referee for a constructive report, and
T. Miyaji, M. Volonteri and L. Ferrarese for useful discussions. 
\end{acknowledgements}

\end{document}